\documentclass[a4paper,11pt]{article}
\usepackage{pifont}
\date{}

\usepackage{makeidx}
\usepackage[centertags]{amsmath}
\usepackage{amsfonts}
\usepackage{amssymb}
\usepackage{amsthm}
\usepackage{epsfig}
\usepackage{color}

\makeatletter \@addtoreset{equation}{section} \makeatother

\begin{document}
\title{\bf Nonplanar Periodic Solutions for Spatial Restricted 3-Body
and 4-Body Problems  \footnote{Supported by National Natural Science
Foundation of China.}}

\author{{ Xiaoxiao Zhao\ and \ Shiqing Zhang}\\
{\small Yangtze Center of Mathematics and College of Mathematics, Sichuan University,}\\
{\small Chengdu 610064, People's Republic of China}} \maketitle
\begin{quote}

{\bf Abstract:} In this paper, we study the existence of non-planar
periodic solutions for the following spatial restricted 3-body and
4-body problems: for $N=2 \ \mbox{or} \ 3$, given any positive
masses $m_{1},\cdots,m_{N}$, the mass points of $m_{1},\cdots,m_{N}$
move in the plane of $N$ circular obits centered at the center of
masses, the sufficiently small mass moves on the perpendicular axis
passing the center of masses. Using variational minimizing methods,
we establish the existence of the minimizers of the Lagrangian
action on anti-T/2 or odd symmetric loop spaces. Moreover, we prove
these minimizers are non-planar periodic solutions by using the
Jacobi's Necessary Condition for local minimizers.


2000 AMS Subject Classification 34C15, 34C25, 58F.
\end{quote}

\section{Introduction and Main Results}
\ \ \ \ \ \ In this paper, we study the spatial circular restricted
3-body and 4-body problems. For $N=2\ \mbox{or}\ 3$, suppose points
of positive masses $m_{1},\cdots,m_{N}$ move in the plane of their
circular orbits $q_{1}(t),\cdots, q_{N}(t)$ with the radius
$r_{1},\cdots,r_{N}>0$ and the center of masses is at the origin;
suppose the sufficiently small mass point does not influence the
motion of $m_{1},\cdots,m_{N}$, and  moves on the vertical axis of
the moving plane for the given masses $m_{1},\cdots,m_{N}$, here the
vertical axis passes through the center of masses. \

It is known that $q_{1}(t),\cdots, q_{N}(t)(N=2 \ \mbox{or} \ 3)$
satisfy the Newtonian equations:
\begin{equation}\label{e1}
m_{i}\ddot{q_{i}}=\frac{\partial U}{\partial q_{i}},\ \ \ \
i=1,\cdots,N,
\end{equation}
where
\begin{equation}\label{e2}
U=\sum\limits_{1\leq i< j\leq N}\frac{m_{i}m_{j} }{| q_{i}-q_{j}|}.
\end{equation}
The orbit $q(t)=(0,0,z(t))\in R^{3}$ for sufficiently small mass is
governed by the gravitational forces of $m_{1},\cdots,m_{N}(N=2 \
\mbox{or} \ 3)$ and therefore it satisfies the following equation
\begin{equation}\label{e3}
\ddot{q}=\sum\limits_{i=1}^{N}\frac{m_{i}(q_{i}-q) }{|
q_{i}-q|^{3}}, \ N=2 \ \mbox{or} \ 3.
\end{equation}
\ \ \ \ \ \ \ For $N\geq2$, there are many papers concerned with
the restricted N-body problem, see [3,4,6,8-10] and the references
therein. In \cite{r11}, Sitnikov considered the following model:
two mass points of equal mass $m_{1} = m_{2} = \frac{1}{2}$ move
in the plane of their elliptic orbits and the center of masses is
at rest, the third mass point which does not influence the motion
of the first two moves on the line perpendicular to the plane
containing the first two mass points and goes through the center
of mass, and he used geometrical methods to prove the existence of
the oscillatory parabolic orbit of
\begin{equation}\label{e5}
\ddot{z}(t)=\frac{-z(t)}{(|z(t)|^{2}+|r(t)|^{2})^{3/2}},
\end{equation}
where $r(t)=r(t+2\pi)>0$ is the distance from the center of mass
to one of the first two mass points. McGehee \cite{r9} used the
stable and unstable manifolds to study the homoclinic orbits
(parabolic orbits) of (\ref{e5}). In\cite{r7}, Mathlouthi studied
the periodic solutions for the spatial circular restricted 3-body
problems by minimax variational methods. Recently, Li, Zhang and
Zhao\cite{r6} used variational minimizing methods to study spatial
circular restricted N+1-body problem with a zero mass moving on
the vertical axis of the moving plane for N equal mass.

\

Motivated by \cite{r6}, we use the Jacobi's Necessary Condition for
local minimizers to further study the spatial circular restricted
3-body and 4-body problems with a sufficiently small mass moving on
the perpendicular axis of the circular orbits plane for any given
masses $m_{1},\cdots,m_{N}(N=2 \ \mbox{or} \ 3)$.

Define
\begin{equation*}
W^{1,2}(R/TZ,R)=\bigg\{u(t)\Big| u(t),u'(t)\in L^{2}(R,R), \
u(t+T)=u(t) \bigg\}.
\end{equation*}
The inner product and the norm of $W^{1,2}(R/TZ,R)$ are
\begin{equation}
<u,v>=\int_{0}^{T}(uv+u'\cdot v')dt,\ \ \ \ \ \ \ \ \ \
\end{equation}
\begin{equation}\label{e28}
\|u\|=\Big[\int_{0}^{T}|u|^{2}dt\Big]^{\frac{1}{2}}+\Big[\int_{0}^{T}
|u'|^{2}dt\Big]^{\frac{1}{2}}.
\end{equation}
The functional corresponding to the equation (\ref{e3}) is
\begin{equation}\label{e4}
\begin{aligned}
f(q)&=\int_{0}^{T}\Big[\frac{1}{2}|\dot{q}|^{2}+\sum\limits_{i=1}^{N}\frac{m_{i}
}{| q-q_{i}|}\Big]dt,\ \ \ \ q\in \Lambda_{j},j=1,2\\
&=\int_{0}^{T}\Big[\frac{1}{2}|z'|^{2}+\frac{m_{1}}{\sqrt{r_{1}^{2}+z^{2}}}+\cdots
+\frac{m_{N}}{\sqrt{r_{N}^{2}+z^{2}}}\Big]dt\triangleq f(z),\ N=2 \
\mbox{or} \ 3,
\end{aligned}
\end{equation}
where
\begin{equation*}
\Lambda_{1}=\left\{ q(t)=(0,0,z(t))\Big|z(t)\in W^{1,2}(R/TZ,R),\
z(t+\frac{T}{2})=-z(t) \right\},
\end{equation*}
and
\begin{equation*}
\Lambda_{2}=\bigg\{ q(t)=(0,0,z(t))\Big|z(t)\in W^{1,2}(R/TZ,R),\
z(-t)=-z(t) \bigg\}.\ \ \ \ \
\end{equation*}

Our main results are the following:

\vspace{0.4cm}\textbf{Theorem 1.1} \ For $N=2$, the minimizer of
$f(q)$ on the closure $\overline{\Lambda}_{i}$ of
$\Lambda_{i}(i=1,2)$ is a nonplanar and noncollision periodic
solution.

\vspace{0.4cm}\textbf{Theorem 1.2} \ For $N=3$, the minimizer of
$f(q)$ on the closure $\overline{\Lambda}_{i}$ of
$\Lambda_{i}(i=1,2)$ is a nonplanar and noncollision periodic
solution.

\section{Preliminaries}
In this section,  we will list some basic Lemmas and inequality for
proving our Theorems 1.1 and 1.2.

\vspace{0.4cm}{\textbf{Lemma 2.1}}(Palais's Symmetry
Principle(\cite{r10}))\ Let $\sigma$ be an orthogonal representation
of a finite or compact group $G$, $H$ be a real Hilbert space,
$f:H\rightarrow R$ satisfies $f(\sigma\cdot x)=f(x),\forall\sigma\in
G,\forall x\in H$.

Set $F=\{x\in H|\sigma\cdot x=x,\ \forall \sigma\in G\}$. Then the
critical point of $f$ in $F$ is also a critical point of $f$ in $H$.

\vspace{0.4cm}{\textbf{Remark 2.1}}\ \ By Palais's Symmetry
Principle, we know that the critical point of $f(q)$ in
$\overline{\Lambda}_{i}=\Lambda_{i}(i=1,2)$ is a periodic solution
of Newtonian equation (\ref{e3}).

\vspace{0.4cm}{\textbf{Lemma 2.2}}(Tonelli\cite{r1}) Let $X$ be a
reflexive Banach space, $S$ be a weakly closed subset of $X$,
$f:S\rightarrow R\cup +\infty$. If $f\not\equiv +\infty$ is weakly
lower semi-continuous and coercive($f(x)\rightarrow +\infty$ as
$\|x\|\rightarrow +\infty$), then $f$ attains its infimum on $S$.

\vspace{0.4cm}\textbf{Lemma 2.3}(Poincare-Wirtinger
Inequality\cite{r8})\ \ Let $q\in W^{1,2}(R/TZ,R^{K})$ and
$\int_{0}^{T}q(t)dt=0$, then
\begin{equation}
\int_{0}^{T}|q(t)|^{2}dt\leq\frac{T^{2}}{4\pi^{2}}\int_{0}^{T}|\dot{q}(t)|^{2}dt.
\end{equation}

\vspace{0.4cm}{\textbf{Lemma 2.4}}\ \ $f(q)$ in (\ref{e4}) attains
its infimum on $\bar{\Lambda}_{i}=\Lambda_{i}(i=1,2)$.

\vspace{0.4cm}{\textbf{Proof.}}\ \ By using Lemma 2.3, for $\forall
z\in \Lambda_{i},\ i=1,2$, the equivalent norm of (\ref{e28}) in
$\Lambda_{i}(i=1,2)$ is
\begin{equation}
\|z\|\cong\Big[\int_{0}^{T}|z'|^{2}dt\Big]^{\frac{1}{2}}.
\end{equation}
Hence by the definitions of $f(q)$, it is easy to see that $f$ is
$C^{1}$ and coercive on $\Lambda_{i}(i=1,2)$. In order to get Lemma
2.4, we only need to prove that $f$ is weakly lower semi-continuous
on $\Lambda_{i}(i=1,2)$. In fact, for $\forall z_{n}\in
\Lambda_{i}$, if $z_{n}\rightharpoonup z$ weakly, by compact
embedding theorem, we have the uniformly convergence:
\begin{equation}
\max\limits_{0\leq t\leq T}|z_{n}(t)-z(t)|\rightarrow 0,\ \ \ \
n\rightarrow\infty,
\end{equation}
which implies
\begin{equation}\label{e24}
\int_{0}^{T}\frac{m_{1}}{\sqrt{r_{1}^{2}+z_{n}^{2}}}+\cdots
+\frac{m_{N}}{\sqrt{r_{N}^{2}+z_{n}^{2}}}dt\rightarrow\int_{0}^{T}
\frac{m_{1}}{\sqrt{r_{1}^{2}+z^{2}}}+\cdots
+\frac{m_{N}}{\sqrt{r_{N}^{2}+z^{2}}}dt,\ N=2 \ \mbox{or} \ 3.
\end{equation}
It is well-known that the norm and its square are weakly lower
semi-continuous. Therefore, combined with (\ref{e24}), one has
\begin{equation*}
\liminf\limits_{n\rightarrow\infty}f(z_{n})\geq f(z),
\end{equation*}
that is, $f$ is weakly lower semi-continuous on
$\Lambda_{i}(i=1,2)$. By lemma 2.2, we can get that $f(q)$ in
(\ref{e4}) attains its infimum on
$\bar{\Lambda}_{i}=\Lambda_{i}(i=1,2)$.\ \ $\Box$

\vspace{0.4cm}{\textbf{Lemma 2.5}}(Jacobi's Necessary
Condition\cite{r4})\ \ Let $F\in C^{3}([a,b]\times R\times R,R)$. If
the critical point $y=\tilde{y}(x)$ corresponds to a minimum of the
functional $\int_{a}^{b}F(x,y(x),y'(x))dx$ on $M=\{y\in
W^{1,2}([a,b],R)|y(a)=A,y(b)=B\}$ and if $F_{y'y'}>0$ along this
critical point, then the open interval $(a,b)$ contains no points
conjugate to $a$, that is, for $\forall c\in(a,b)$, the following
problem:
\begin{equation}
\label{} \left\{\begin{array}{ll}
-\frac{d}{dx}(Ph')+Qh=0,& \\
h(a)=0,\ \ h(c)=0,
\end{array}\right.
\end{equation}
has only the trivial solution $h(x)\equiv0,\ \forall x\in(a,c)$,
where
\begin{equation}
P=\frac{1}{2}F_{y'y'}|_{y=\tilde{y}},\ \ \ \ \ \ \ \ \ \ \ \ \ \ \
\end{equation}
\begin{equation}
Q=\frac{1}{2}\Big(F_{yy}-\frac{d}{dx}F_{yy'}\Big)\Big|_{y=\tilde{y}}.
\end{equation}

\vspace{0.4cm}{\textbf{Remark 2.2}}\ \ It is easy to see that Lemma
2.5 is suitable for the fixed end problem. In this paper, we
consider the periodic solutions of (\ref{e3}) on
$\overline{\Lambda}_{i}=\Lambda_{i}(i=1,2)$, hence we need to
establish a similar conclusion as Lemma 2.5 for the periodic
boundary problem.

\vspace{0.4cm}{\textbf{Lemma 2.6}}\ \ Let $F\in C^{3}(R\times
R\times R,R)$. Assume that $u=\tilde{u}(t)$ is a critical point of
the functional $\int_{0}^{T}F(t,u(t),u'(t))dt$ on $W^{1,2}(R/TZ,R)$
and $F_{u'u'}|_{u=\tilde{u}}>0$. If the open interval $(0,T)$
contains a point $c$ conjugate to $0$, then $u=\tilde{u}(t)$ is not
a minimum of the functional $\int_{0}^{T}F(t,u(t),u'(t))dt$.

\vspace{0.4cm}{\textbf{Proof.}}\ \ Suppose $u=\tilde{u}(t)$ is a
minimum of the functional $\int_{0}^{T}F(t,u(t),u'(t))dt$. The
second variation of $\int_{0}^{T}F(t,u(t),u'(t))dt$ is
\begin{equation}
\int_{0}^{T}(Ph'^{2}+Qh^{2})dt,\ \ \ \ \ \ \ \ \ \ \ \ \ \ \ \ \ \
\end{equation}
where
\begin{equation}
P=\frac{1}{2}F_{u'u'}|_{u=\tilde{u}},\ \ \ \ \ \ \ \ \ \ \ \ \ \ \
\end{equation}
\begin{equation}
Q=\frac{1}{2}\Big(F_{uu}-\frac{d}{dt}F_{uu'}\Big)\Big|_{u=\tilde{u}}.
\end{equation}
Set
\begin{equation}\label{e25}
Q_{\tilde{u}}(h)=\int_{0}^{T}(Ph'^{2}+Qh^{2})dt.
\end{equation}
For $\forall h\in C_{0}^{1}([0,T],R)$, it is easy to see that
$Q_{\tilde{u}}(h)\geq0$. Then by $Q_{\tilde{u}}(\theta)=0$, $\theta$
is a minimum of $Q_{\tilde{u}}(h)$. The Euler-Lagrange equation
which is called the Jacobi equation of (\ref{e25}) is
\begin{equation}
-\frac{d}{dt}(Ph')+Qh=0.\ \ \ \ \ \ \ \ \ \ \ \ \ \ \ \ \ \
\end{equation}
Since the interval $(0,T)$ contains a point $c$ conjugate to $0$,
there exists a nonzero Jacobi field $h_{0}\in C^{2}([0,T],R)$
satisfying
\begin{equation}
\left\{\begin{array}{ll}
-\frac{d}{dt}(Ph_{0}')+Qh_{0}=0,& \\
h_{0}(0)=0,\ \ h_{0}(c)=0.
\end{array}\right.
\end{equation}
Let
\begin{equation}
\hat{h}(t)=\left\{\begin{array}{ll}
h_{0}(t)& \ \ \ \ t\in[0,c],\\
0&\ \ \ \ t\in(c,T],
\end{array}\right.
\end{equation}
we have $\hat{h}\in C^{2}([0,T]\backslash\{c\},R)$,
$\hat{h}(0)=\hat{h}(c)=\hat{h}(T)=0$ and
\begin{equation}\label{e30}
Q_{\tilde{u}}(\hat{h})=\int_{0}^{T}(P\hat{h}'^{2}+Q\hat{h}^{2})dt
=\int_{0}^{c}(Ph_{0}'^{2}+Qh_{0}^{2})dt=0.
\end{equation}
Notice that we can extend $\hat{h}$ periodically when we take T as
the period, so $\hat{h}\in W_{0}^{1,2}(R/TZ,R)$. For $\forall h\in
C_{0}^{1}([0,T],R)$, it is easy to check that
$Q_{\tilde{u}}(h)\geq0$. Then by (\ref{e30}), one has $\hat{h}\in
C^{2}([0,T]\backslash\{c\},R)\cap W_{0}^{1,2}(R/TZ,R)$ is a
minimum of $Q_{\tilde{u}}(h)$. Hence we get
\begin{equation}
-\frac{d}{dt}(P\hat{h}')+Q\hat{h}=0.\ \ \ \ \ \ \ \ \ \ \ \ \ \ \ \
\ \
\end{equation}
Combine with $\hat{h}(0)=\hat{h}(c)=0$, by the uniqueness of initial
value problems for second order differential equation, we have
$\hat{h}(t)\equiv0$ on [0,c], which contradicts the definition of
$\hat{h}$. Therefore, Lemma 2.6 holds. $\Box$

\section{Proof of Theorem 1.1}
In this section, we consider the spatial circular restricted 3-body
problem with a sufficiently small mass moving on the vertical axis
of the moving plane for arbitrary given positive masses
$m_{1},m_{2}$. Suppose the planar circular orbits are
\begin{equation}\label{e6}
q_{1}(t)=r_{1}e^{\sqrt{-1}\frac{2\pi}{T} t}, \ \ \ \
q_{2}(t)=-r_{2}e^{\sqrt{-1}\frac{2\pi}{T} t},
\end{equation}
here the radius $r_{1},r_{2}$ are positive constants depending on
$m_{i}(i=1,2)$ and $T$ (see Lemma 3.1). We also assume that
\begin{equation}\label{e7}
m_{1}q_{1}(t)+m_{2} q_{2}(t)=0.
\end{equation}
The functional corresponding to the equation (\ref{e3}) is
\begin{equation}\label{e8}
\begin{aligned}
f(q)&=\int_{0}^{T}\Big[\frac{1}{2}|\dot{q}|^{2}+\frac{m_{1}}{|
q-q_{1}|}+\frac{m_{2}}{|
q-q_{2}|}\Big]dt,\ \ \ \ q\in \Lambda_{i},\ i=1,2\\
&=\int_{0}^{T}\Big[\frac{1}{2}|z'|^{2}+\frac{m_{1}}{\sqrt{r_{1}^{2}+z^{2}}}
+\frac{m_{2}}{\sqrt{r_{2}^{2}+z^{2}}}\Big]dt\triangleq f(z).
\end{aligned}
\end{equation}

\vspace{0.4cm}{\textbf{Lemma 3.1}}\ \ The radius $r_{1},r_{2}$ of
the planar circular orbits for the masses $m_{1},m_{2}$ are
\begin{equation*}
r_{1}=\Big(\frac{T}{2\pi(m_{1}+m_{2})}\Big)^{\frac{2}{3}}m_{2}, \ \
\ \ r_{2}=\Big(\frac{T}{2\pi(m_{1}+m_{2})}\Big)^{\frac{2}{3}}m_{1}.
\end{equation*}

\vspace{0.4cm}{\textbf{Proof.}}\ \ Substituting (\ref{e6}) into
(\ref{e7}), it is easy to get
\begin{equation}\label{e29}
r_{2}=\frac{m_{1}}{m_{2}}r_{1}.
\end{equation}
It follows from (\ref{e1}) and
(\ref{e2}) that
\begin{equation}
\ddot{q_{1}}=m_{2}\frac{q_{2}-q_{1}}{| q_{2}-q_{1}|^{3}}.
\end{equation}
Then by (\ref{e6}) and (\ref{e29}), we have
\begin{equation}
-\frac{4\pi^{2}}{T^{2}}q_{1}=m_{2}\frac{(-\frac{m_{1}}{m_{2}}-1)q_{1}}{r_{1}^{3}|
-\frac{m_{1}}{m_{2}}-1|^{3}},
\end{equation}
which implies
\begin{equation}\label{e9}
r_{1}=\Big(\frac{T}{2\pi(m_{1}+m_{2})}\Big)^{\frac{2}{3}}m_{2}.\ \ \
\ \
\end{equation}
Hence by (\ref{e29}), one has
\begin{equation}\label{e10}
r_{2}=\Big(\frac{T}{2\pi(m_{1}+m_{2})}\Big)^{\frac{2}{3}}m_{1}. \ \
\ \Box
\end{equation}

\vspace{0.4cm}\textbf{Proof of Theorem 1.1}\ \ Clearly,
$q(t)=(0,0,0)$ is a critical point of $f(q)$ on
$\bar{\Lambda}_{i}=\Lambda_{i}(i=1,2)$. For the functional
(\ref{e8}), let
\begin{equation*}
F(z,z')=\frac{1}{2}|z'|^{2}+\frac{m_{1}}{\sqrt{r_{1}^{2}+z^{2}}}+\frac{m_{2}}{\sqrt{r_{2}^{2}+z^{2}}}.
\end{equation*}
Then the second variation of (\ref{e8}) in the neighborhood of $z=0$
is given by
\begin{equation}\label{e11}
\int_{0}^{T}(Ph'^{2}+Qh^{2})dt,\ \ \ \ \ \ \ \ \ \ \ \ \ \ \ \ \ \
\end{equation}
where
\begin{equation}
P=\frac{1}{2}F_{z'z'}|_{z=0}=\frac{1}{2},\ \ \ \ \ \ \ \ \ \ \ \ \ \
\ \ \ \ \ \ \ \ \ \ \ \ \ \ \ \ \ \
\end{equation}
\begin{equation}
Q=\frac{1}{2}\Big(F_{zz}-\frac{d}{dt}F_{zz'}\Big)\Big|_{z=0}=-\Big(\frac{m_{1}}{2r_{1}^{3}}+\frac{m_{2}}{2r_{2}^{3}}\Big).
\end{equation}
The Euler equation of (\ref{e11}) is called the Jacobi equation of
the original functional (\ref{e8}), which is
\begin{equation}\label{e22}
-\frac{d}{dt}(Ph')+Qh=0,\ \ \ \ \ \ \ \ \ \ \ \ \
\end{equation}
that is,
\begin{equation}\label{e12}
h''+\Big(\frac{m_{1}}{r_{1}^{3}}+\frac{m_{2}}{r_{2}^{3}}\Big)h=0.\ \
\ \ \ \ \ \ \ \ \ \ \
\end{equation}
Next, we study the solution of (\ref{e12}) with initial values
$h(0)=0,\ h'(0)=1$. It is easy to get
\begin{equation}
h(t)=\sqrt{\frac{r_{1}^{3}r_{2}^{3}}{m_{2}r_{1}^{3}+m_{1}r_{2}^{3}}}\cdot
sin\sqrt{\frac{m_{1}}{r_{1}^{3}}+\frac{m_{2}}{r_{2}^{3}}}t.
\end{equation}
It follows from (\ref{e9}) and (\ref{e10})that
\begin{equation}
\sqrt{\frac{m_{1}}{r_{1}^{3}}+\frac{m_{2}}{r_{2}^{3}}}=
\sqrt{\frac{m_{1}^{4}+m_{2}^{4}}{m_{1}^{3}m_{2}^{3}}}(m_{1}+m_{2})\cdot\frac{2\pi}{T}.
\end{equation}
Hence
\begin{equation}
h(t)=\frac{\sqrt{m_{1}^{3}m_{2}^{3}}T}{2\pi\sqrt{m_{1}^{4}+m_{2}^{4}}(m_{1}+m_{2})}
\cdot
sin\Bigg(\sqrt{\frac{m_{1}^{4}+m_{2}^{4}}{m_{1}^{3}m_{2}^{3}}}(m_{1}+m_{2})
\cdot\frac{2\pi}{T}t\Bigg),
\end{equation}
which is not identically zero on
$[0,\frac{\sqrt{m_{1}^{3}m_{2}^{3}}T}{\sqrt{m_{1}^{4}+m_{2}^{4}}(m_{1}+m_{2})}]$.
Since
\begin{equation}
m_{1}^{6}+m_{2}^{6}\geq2\sqrt{m_{1}^{6}\cdot
m_{2}^{6}}=2m_{1}^{3}m_{2}^{3}> m_{1}^{3}m_{2}^{3},
\end{equation}
one has
\begin{equation}
\begin{aligned}
(m_{1}^{4}+m_{2}^{4})(m_{1}+m_{2})^{2}&>m_{1}^{6}+m_{2}^{6}\\
&> m_{1}^{3}m_{2}^{3},
\end{aligned}
\end{equation}
which implies
\begin{equation}
\frac{\sqrt{m_{1}^{3}m_{2}^{3}}}{\sqrt{m_{1}^{4}+m_{2}^{4}}(m_{1}+m_{2})}<1.
\end{equation}
Therefore
\begin{equation}
\frac{\sqrt{m_{1}^{3}m_{2}^{3}}T}{\sqrt{m_{1}^{4}+m_{2}^{4}}(m_{1}+m_{2})}<T.
\end{equation}
Choose
$0<c=\frac{\sqrt{m_{1}^{3}m_{2}^{3}}T}{2\sqrt{m_{1}^{4}+m_{2}^{4}}(m_{1}+m_{2})}
<\frac{T}{2}$, we have
\begin{equation}
h(c)=\frac{\sqrt{m_{1}^{3}m_{2}^{3}}T}{2\pi\sqrt{m_{1}^{4}+m_{2}^{4}}(m_{1}+m_{2})}
\cdot sin\pi=0.
\end{equation}

\

Case 1: Minimizing $f(q)$ on $\bar{\Lambda}_{1}=\Lambda_{1}$.

\

Let
\begin{equation}
\tilde{h}(t)=\left\{\begin{array}{ll}
h(t)& \ \ \ \ t\in[0,c],\\
0& \ \ \ \ t\in(c,\frac{T}{2}],\\
-h(t-\frac{T}{2})& \ \ \ \ t\in(\frac{T}{2},\frac{T}{2}+c],\\
0& \ \ \ \ t\in(\frac{T}{2}+c,T].
\end{array}\right.
\end{equation}
It is easy to check that $\tilde{h}(t)\in
C^{2}([0,T]\backslash\{c,\frac{T}{2},\frac{T}{2}+c\},R)\cap
W^{1,2}(R,R)$, $\tilde{h}(t+\frac{T}{2})=-\tilde{h}(t)$,
$\tilde{h}(0)=h(0)=0$, $\tilde{h}(c)=h(c)=0$ and $\tilde{h}$ is a
nonzero solution of (\ref{e22}). Notice that we can extend
$\tilde{h}$ periodically when we take T as the period, so
$\tilde{h}\in\Lambda_{1}$. Then by Lemma 2.6, $q(t)=(0,0,0)$ is
not a local minimum for $f(q)$ on $\Lambda_{1}$. Hence the
minimizers of $f(q)$ on $\Lambda_{1}$ are not always at the center
of masses, they must oscillate periodically on the vertical axis,
that is, the minimizers are not always co-planar, therefore, we
get the non-planar periodic solutions.

\

Case 2: Minimizing $f(q)$ on $\bar{\Lambda}_{2}=\Lambda_{2}$.

\

Let
\begin{equation}
\bar{h}(t)=\left\{\begin{array}{ll}
h(t)& \ \ \ \ t\in[0,c],\\
0& \ \ \ \ t\in(c,T-c],\\
-h(T-t)& \ \ \ \ t\in(T-c,T].
\end{array}\right.
\end{equation}
It is easy to check that $\bar{h}(t)\in
C^{2}([0,T]\backslash\{c,T-c\},R)\cap W^{1,2}(R,R)$,
$\bar{h}(-t)=-\bar{h}(t)$, $\bar{h}(0)=h(0)=0$,
$\bar{h}(c)=h(c)=0$ and $\bar{h}$ is a nonzero solution of
(\ref{e22}). Notice that we can extend $\bar{h}$ periodically when
we take T as the period, so $\bar{h}\in\Lambda_{2}$. Then by Lemma
2.6, $q(t)=(0,0,0)$ is not a local minimum for $f(q)$ on
$\Lambda_{2}$. Hence the minimizers of $f(q)$ on $\Lambda_{2}$ are
not always at the center of masses, they must oscillate
periodically on the vertical axis, that is, the minimizers are not
always co-planar, therefore, we get the non-planar periodic
solutions.

\

Combined with Lemma 2.4, the proof is completed.\ \ $\Box$

\section{Proof of Theorem 1.2}
In this section, we consider the spatial circular restricted 4-body
problem with a sufficiently small mass moving on the vertical axis
of the moving plane for arbitrary positive masses
$m_{1},m_{2},m_{3}$. Suppose there exists
$\theta_{1},\theta_{2},\theta_{3}\in(0,2\pi)$ such that the planar
circular orbits are
\begin{equation}\label{e13}
q_{1}(t)=r_{1}e^{\sqrt{-1}\frac{2\pi}{T} t}e^{\sqrt{-1}\theta_{1}},\
\ q_{2}(t)=r_{2}e^{\sqrt{-1}\frac{2\pi}{T}
t}e^{\sqrt{-1}\theta_{2}}, \ \
q_{3}(t)=r_{3}e^{\sqrt{-1}\frac{2\pi}{T} t}e^{\sqrt{-1}\theta_{3}},
\end{equation}
here the radius $r_{1}, r_{2},r_{3}$ are positive constants
depending on $m_{i}(i=1,2,3)$ and $T$ (see Lemma 4.2). We also
assume that
\begin{equation}\label{e14}
m_{1}q_{1}(t)+m_{2} q_{2}(t)+m_{3} q_{3}(t)=0
\end{equation}
and
\begin{equation}\label{e15}
|q_{i}-q_{j}|=l, \ \ 1\leq i\neq j\leq3,
\end{equation}
where the constant $l>0$ depends on $m_{i}(i=1,2,3)$ and $T$ (see
Lemma 4.1). The functional corresponding to the equation (\ref{e3})
is
\begin{equation}\label{e16}
\begin{aligned}
f(q)&=\int_{0}^{T}\Big[\frac{1}{2}|\dot{q}|^{2}+\frac{m_{1}}{|
q-q_{1}|}+\frac{m_{2}}{| q-q_{2}|}+\frac{m_{3}}{| q-q_{3}|}\Big]dt,\
\ \ \ q\in
\Lambda_{i},\ i=1,2\\
&=\int_{0}^{T}\Big[\frac{1}{2}|z'|^{2}+\frac{m_{1}}{\sqrt{r_{1}^{2}+z^{2}}}
+\frac{m_{2}}{\sqrt{r_{2}^{2}+z^{2}}}++\frac{m_{3}}{\sqrt{r_{3}^{2}+z^{2}}}\Big]dt\triangleq
f(z).
\end{aligned}
\end{equation}
In order to get Theorem 1.2, we firstly prove Lemmas 4.1 and 4.2
as follows.

\vspace{0.4cm}{\textbf{Lemma 4.1}}\ \ Let $M=m_{1}+m_{2}+m_{3}$, we
have $l=\sqrt[3]{\frac{MT^{2}}{4\pi^{2}}}$.

\vspace{0.4cm}{\textbf{Proof.}}\ \  It follows from (\ref{e1}) and
(\ref{e2}) that
\begin{equation}
\ddot{q_{1}}=m_{2}\frac{q_{2}-q_{1}}{|
q_{2}-q_{1}|^{3}}+m_{3}\frac{q_{3}-q_{1}}{| q_{3}-q_{1}|^{3}}.
\end{equation}
Then by (\ref{e13})-(\ref{e15}), one has
\begin{equation}
\begin{aligned}
-\frac{4\pi^{2}}{T^{2}}q_{1}&=\frac{1}{l^{3}}(m_{2}q_{2}+m_{3}q_{3}-m_{2}q_{1}-m_{3}q_{1})\\
&=\frac{1}{l^{3}}(-m_{1}q_{1}-m_{2}q_{1}-m_{3}q_{1}),
\end{aligned}
\end{equation}
which implies
\begin{equation}
l^{3}=\frac{MT^{2}}{4\pi^{2}},\ \ \ \ \ \
\end{equation}
that is,
\begin{equation}\label{e17}
l=\sqrt[3]{\frac{MT^{2}}{4\pi^{2}}}.\ \ \ \Box
\end{equation}

\vspace{0.4cm}{\textbf{Lemma 4.2}}\ \ The radius $r_{1},r_{2},r_{3}$
of the planar circular orbits for the masses $m_{1},m_{2},m_{3}$ are
\begin{equation*}
r_{1}=\frac{\sqrt{m^{2}_{2}+m_{2}m_{3}+m^{2}_{3}}}{M}l,\ \ \ \ \ \
\end{equation*}
\begin{equation*}
r_{2}=\frac{\sqrt{m^{2}_{1}+m_{1}m_{3}+m^{2}_{3}}}{M}l,\ \ \ \ \ \
\end{equation*}
\begin{equation*}
r_{3}=\frac{\sqrt{m^{2}_{1}+m_{1}m_{2}+m^{2}_{2}}}{M}l.\ \ \ \ \ \
\end{equation*}

\vspace{0.4cm}{\textbf{Proof.}}\ \  Choose the geometrical center of
the initial configuration ($q_{1}(0),q_{2}(0),q_{3}(0)$) as the
origin of the coordinate (x,y). Without loss of generality, by
(\ref{e15}), we suppose the location coordinates of
$q_{1}(0),q_{2}(0),q_{3}(0)$ are
$A_{1}(\frac{\sqrt{3}l}{3},0),A_{2}(-\frac{\sqrt{3}l}{6},\frac{l}{2}),
A_{3}(-\frac{\sqrt{3}l}{6},-\frac{l}{2})$. Then we can get the
coordinate of the center of masses $m_{1},m_{2},m_{3}$ is
$C(\frac{\frac{\sqrt{3}}{3}m_{1}l-\frac{\sqrt{3}}{6}m_{2}l
-\frac{\sqrt{3}}{6}m_{3}l}{M},\frac{\frac{m_{2}}{2}l-\frac{m_{3}}{2}l}{M})$.
To make sure the Assumption (\ref{e14}) holds, we introduce the new
coordinate
\begin{equation*}
\left\{\begin{array}{ll} X=x-
\frac{\frac{\sqrt{3}}{3}m_{1}l-\frac{\sqrt{3}}{6}m_{2}l
-\frac{\sqrt{3}}{6}m_{3}l}{M},\\
Y=y-\frac{\frac{m_{2}}{2}l-\frac{m_{3}}{2}l}{M} .
\end{array}\right.
\end{equation*}
Hence in the new coordinate (X,Y), the location coordinates of
$q_{1}(0),q_{2}(0),q_{3}(0)$ are
$A_{1}(\frac{\frac{\sqrt{3}}{2}m_{2}l+\frac{\sqrt{3}}{2}m_{3}l}{M},$
$\frac{-\frac{m_{2}}{2}l+\frac{m_{3}}{2}l}{M}),$
$A_{2}(-\frac{\frac{\sqrt{3}}{2}m_{1}l}{M},\frac{\frac{m_{1}}{2}l+m_{3}l}{M}),
A_{3}(-\frac{\frac{\sqrt{3}}{2}m_{1}l}{M},-\frac{\frac{m_{1}}{2}l+m_{2}l}{M})$
and the center of masses $m_{1},m_{2},m_{3}$ is at the origin
$O(0,0)$. Then compared with (\ref{e13}), we have
\begin{equation}\label{e18}
r_{1}=|A_{1}O|=\frac{\sqrt{m^{2}_{2}+m_{2}m_{3}+m^{2}_{3}}}{M}l,\ \
\ \ \ \
\end{equation}
\begin{equation}\label{e19}
r_{2}=|A_{2}O|=\frac{\sqrt{m^{2}_{1}+m_{1}m_{3}+m^{2}_{3}}}{M}l,\ \
\ \ \ \
\end{equation}
\begin{equation}\label{e20}
r_{3}=|A_{3}O|=\frac{\sqrt{m^{2}_{1}+m_{1}m_{2}+m^{2}_{2}}}{M}l,\ \
\ \ \ \
\end{equation}
and
\begin{equation}
\tan\theta_{1}=\frac{\sqrt{3}(-m_{2}+m_{3})}{3(m_{2}+m_{3})},\ \ \ \
\tan\theta_{2}=-\frac{\sqrt{3}(m_{1}+2m_{3})}{3m_{1}},\ \ \ \
\tan\theta_{3}=\frac{\sqrt{3}(m_{1}+2m_{2})}{3m_{1}}.\ \ \ \Box
\end{equation}

\vspace{0.4cm}\textbf{Proof of Theorem 1.2}\ \ Clearly,
$q(t)=(0,0,0)$ is a critical point of $f(q)$ on
$\bar{\Lambda}_{i}=\Lambda_{i}(i=1,2)$. For the functional
(\ref{e16}), let
\begin{equation*}
F(z,z')=\frac{1}{2}|z'|^{2}+\frac{m_{1}}{\sqrt{r_{1}^{2}+z^{2}}}
+\frac{m_{2}}{\sqrt{r_{2}^{2}+z^{2}}}+\frac{m_{3}}{\sqrt{r_{3}^{2}+z^{3}}}.
\end{equation*}
Then the second variation of (\ref{e16}) in the neighborhood of
$z=0$ is given by
\begin{equation}\label{e26}
\int_{0}^{T}(Ph'^{2}+Qh^{2})dt,\ \ \ \ \ \ \ \ \ \ \ \ \ \ \ \ \ \
\end{equation}
where
\begin{equation}
P=\frac{1}{2}F_{z'z'}|_{z=0}=1,\ \ \ \ \ \ \ \ \ \ \ \ \ \ \ \ \ \ \
\ \ \ \ \ \ \ \ \ \ \ \ \ \ \ \ \ \ \ \ \ \
\end{equation}
\begin{equation}
Q=\frac{1}{2}\Big(F_{zz}-\frac{d}{dt}F_{zz'}\Big)\Big|_{z=0}=-\Big(\frac{m_{1}}{2r_{1}^{3}}
+\frac{m_{2}}{2r_{2}^{3}}+\frac{m_{3}}{2r_{3}^{3}}\Big).
\end{equation}
The Euler equation of (\ref{e26}) is called the Jacobi equation of
the original functional (\ref{e16}), which is
\begin{equation}\label{e23}
-\frac{d}{dt}(Ph')+Qh=0,\ \ \ \ \ \ \ \ \ \ \ \ \
\end{equation}
that is,
\begin{equation}\label{e27}
h''+\Big(\frac{m_{1}}{r_{1}^{3}}+\frac{m_{2}}{r_{2}^{3}}+\frac{m_{3}}{r_{3}^{3}}\Big)h=0.\
\ \ \ \ \ \ \ \ \ \ \ \
\end{equation}
Next, we study the solution of (\ref{e27}) with initial values
$h(0)=0,\ h'(0)=1$. It is easy to get
\begin{equation}
h(t)=\sqrt{\frac{r_{1}^{3}r_{2}^{3}r_{3}^{3}}{m_{3}r_{1}^{3}r_{2}^{3}
+m_{2}r_{1}^{3}r_{3}^{3}+m_{1}r_{2}^{3}r_{3}^{3}}}\cdot
sin\sqrt{\frac{m_{1}}{r_{1}^{3}}+\frac{m_{2}}{r_{2}^{3}}+\frac{m_{3}}{r_{3}^{3}}}t.
\end{equation}
Let
\begin{equation*}
A=\frac{\sqrt{m^{2}_{2}+m_{2}m_{3}+m^{2}_{3}}}{M},\ \ \ \ \ \ \ \ \
\end{equation*}
\begin{equation*}
B=\frac{\sqrt{m^{2}_{1}+m_{1}m_{3}+m^{2}_{3}}}{M},\ \ \ \ \ \ \ \ \
\end{equation*}
\begin{equation*}
C=\frac{\sqrt{m^{2}_{1}+m_{1}m_{2}+m^{2}_{2}}}{M}.\ \ \ \ \ \ \ \ \
\end{equation*}
It follows from (\ref{e17})-(\ref{e20}) that
\begin{equation}
\begin{aligned}
\sqrt{\frac{m_{1}}{r_{1}^{3}}+\frac{m_{2}}{r_{2}^{3}}+\frac{m_{3}}{r_{3}^{3}}}
&=\sqrt{\frac{m_{1}}{A^{3}}+\frac{m_{2}}{B^{3}}+\frac{m_{3}}{C^{3}}}\sqrt{\frac{1}{l^{3}}}\\
&=\sqrt{\frac{m_{1}}{A^{3}}+\frac{m_{2}}{B^{3}}+\frac{m_{3}}{C^{3}}}\cdot\frac{2\pi}{\sqrt{M}T}.
\end{aligned}
\end{equation}
Hence
\begin{equation}
h(t)=\frac{\sqrt{M}T}{2\pi\sqrt{\frac{m_{1}}{A^{3}}+\frac{m_{2}}{B^{3}}+\frac{m_{3}}{C^{3}}}}\cdot
sin\Bigg(\sqrt{\frac{m_{1}}{A^{3}}+\frac{m_{2}}{B^{3}}+\frac{m_{3}}{C^{3}}}\cdot\frac{2\pi}{\sqrt{M}T}t\Bigg),
\end{equation}
which is not identically zero on
$[0,\frac{\sqrt{M}T}{\sqrt{\frac{m_{1}}{A^{3}}+\frac{m_{2}}{B^{3}}+\frac{m_{3}}{C^{3}}}}]$.
It is easy to check that
\begin{equation}
\begin{aligned}
M^{2}&>m^{2}_{1}+m_{1}m_{2}+m^{2}_{2},\ \ \ \ \ \ \ \ \ \\
M^{2}&>m^{2}_{1}+m_{1}m_{3}+m^{2}_{3},\ \ \ \ \ \ \ \ \ \\
M^{2}&>m^{2}_{2}+m_{2}m_{3}+m^{2}_{3},\ \ \ \ \ \ \ \ \ \\
\end{aligned}
\end{equation}
which implies
\begin{equation}
\frac{m_{1}}{A^{3}}+\frac{m_{2}}{B^{3}}+\frac{m_{3}}{C^{3}}>m_{1}+m_{2}+m_{3}
=M.
\end{equation}
Therefore
\begin{equation}\label{e21}
\frac{\sqrt{M}T}{\sqrt{\frac{m_{1}}{A^{3}}+\frac{m_{2}}{B^{3}}+\frac{m_{3}}{C^{3}}}}<T.
\end{equation}
Choose
$0<c=\frac{\sqrt{M}T}{2\sqrt{\frac{m_{1}}{A^{3}}+\frac{m_{2}}{B^{3}}
+\frac{m_{3}}{C^{3}}}}<\frac{T}{2}$, we have
\begin{equation}
h(c)=\frac{\sqrt{M}T}{2\pi\sqrt{\frac{m_{1}}{A^{3}}+\frac{m_{2}}{B^{3}}+\frac{m_{3}}{C^{3}}}}\cdot
sin\pi=0.
\end{equation}

\

Case 1: Minimizing $f(q)$ on $\bar{\Lambda}_{1}=\Lambda_{1}$.

\

Let
\begin{equation}
\tilde{h}(t)=\left\{\begin{array}{ll}
h(t)& \ \ \ \ t\in[0,c],\\
0& \ \ \ \ t\in(c,\frac{T}{2}],\\
-h(t-\frac{T}{2})& \ \ \ \ t\in(\frac{T}{2},\frac{T}{2}+c],\\
0& \ \ \ \ t\in(\frac{T}{2}+c,T].
\end{array}\right.
\end{equation}
It is easy to check that $\tilde{h}(t)\in
C^{2}([0,T]\backslash\{c,\frac{T}{2},\frac{T}{2}+c\},R)\cap
W^{1,2}(R,R)$, $\tilde{h}(t+\frac{T}{2})=-\tilde{h}(t)$,
$\tilde{h}(0)=h(0)=0$, $\tilde{h}(c)=h(c)=0$ and $\tilde{h}$ is a
nonzero solution of (\ref{e23}). Notice that we can extend
$\tilde{h}$ periodically when we take T as the period, so
$\tilde{h}\in\Lambda_{1}$. Then by Lemma 2.6, $q(t)=(0,0,0)$ is
not a local minimum for $f(q)$ on $\Lambda_{1}$. Hence the
minimizers of $f(q)$ on $\Lambda_{1}$ are not always at the center
of masses, they must oscillate periodically on the vertical axis,
that is, the minimizers are not always co-planar, therefore, we
get the non-planar periodic solutions.

\

Case 2: Minimizing $f(q)$ on $\bar{\Lambda}_{2}=\Lambda_{2}$.

\

Let
\begin{equation}
\bar{h}(t)=\left\{\begin{array}{ll}
h(t)& \ \ \ \ t\in[0,c],\\
0& \ \ \ \ t\in(c,T-c],\\
-h(T-t)& \ \ \ \ t\in(T-c,T].
\end{array}\right.
\end{equation}
It is easy to check that $\bar{h}(t)\in
C^{2}([0,T]\backslash\{c,T-c\},R)\cap W^{1,2}(R,R)$,
$\bar{h}(-t)=-\bar{h}(t)$, $\bar{h}(0)=h(0)=0$,
$\bar{h}(c)=h(c)=0$ and $\bar{h}$ is a nonzero solution of
(\ref{e23}). Notice that we can extend $\bar{h}$ periodically when
we take T as the period, so $\bar{h}\in\Lambda_{2}$. Then by Lemma
2.6, $q(t)=(0,0,0)$ is not a local minimum for $f(q)$ on
$\Lambda_{2}$. Hence the minimizers of $f(q)$ on $\Lambda_{2}$ are
not always at the center of masses, they must oscillate
periodically on the vertical axis, that is, the minimizers are not
always co-planar, therefore, we get the non-planar periodic
solutions.

\

Combined with Lemma 2.4, the proof is completed.\ \ $\Box$

\end{document}